\documentclass[prl,twocolumn,showpacs,amsmath,amssymb]{revtex4}
\usepackage{graphicx}
\usepackage{bm}
\begin{document}
\title{Ehrenfest time dependent suppression of weak localization}
\author{\.I. Adagideli}
\affiliation{Instituut-Lorentz, Universiteit Leiden, P.O. Box
9506, 2300 RA Leiden, The Netherlands}
\date{\today}
\begin{abstract}
The Ehrenfest time dependence of the suppression of the weak localization
correction to the conductance of a {\em clean} chaotic
cavity is calculated. Unlike in earlier work,
%by Aleiner and Larkin~\cite{REF:Aleiner96},
no impurity scattering is invoked to
imitate diffraction effects. The calculation extends the semiclassical
theory of K. Richter and M. Sieber [Phys. Rev. Lett. {\bf 89}, 206801 (2002)] to include the effect
of a finite Ehrenfest time.
\end{abstract}
\pacs{73.23.-b, 05.45.Mt, 73.20.Fz \\[-7mm]}
%72.15.Rn Localization effects (Anderson or weak localization)
%73.20.Fz Weak or Anderson localization
%73.23.-b Electronic transport in mesoscopic systems
%05.45.Mt Quantum chaos; semiclassical methods
\maketitle

The average conductivity of a disordered metal is reduced with respect to the classical
value by quantum interference.
This phenomenon, known as weak localization, has
been understood long ago~\cite{REF:revWL,REF:weakloc1,REF:weakloc2} in terms of the constructive
interference of time-reversed diffusive trajectories.
Weak localization exists also in quantum dots, which are so small and clean that impurity scattering
can be neglected~\cite{REF:mwl}. In such
ballistic cavities, quantum interference effects develop only after a time scale on which
a minimal wave packet has spread to cover the entire cavity. This time scale, known as
the Ehrenfest time~\cite{REF:Zaslavsky}, is of order $\tau_{\rm E}=\lambda^{-1}\ln k_{F} L$,
with $\lambda$ the Lyapunov exponent of the chaotic dynamics, $k_{F}$ the Fermi wavevector,
and $L$ the linear size of the cavity. The time scale $\tau_{\rm E}$ becomes important if it
is larger than the mean dwell time $\tau_{\rm D}$ of an electron in the quantum dot, coupled
via two point contacts to electron reservoirs.

Suppression of weak localization in the Ehrenfest regime $\tau_{\rm D}<\tau_{\rm E}$ was
first proposed and studied by Aleiner and
Larkin~\cite{REF:Aleiner96}. Their calculation played a seminal role in the development
of the subject, but it was unsatisfactory in one key aspect: A small amount of impurity
scattering was introduced by hand
to imitate the effects of diffraction in a ballistic system. The main aim of our work
is to provide a derivation of the weak
localization correction in the Ehrenfest regime without recourse to impurity
scattering.  To our knowledge no such derivation exists.

The theoretical framework that we shall adopt is the semiclassical
theory of Richter and Sieber~\cite{REF:Sieber}, which is a well-understood
and controlled approximation scheme. In Ref.~\cite{REF:Sieber} the effects
of finite $\tau_{\rm E}$ were not considered, so there the weak localization
correction was given by the value known from random matrix
theory~\cite{REF:Jalabert,REF:Baranger}. We find that the absence of interfering
trajectories when $\tau_{\rm D}<\tau_{\rm E}$ leads to the exponential suppression of
the weak localization correction $\propto \exp(-\tau_{\rm E}/\tau_{\rm D})$, in
agreement with Ref.~\cite{REF:Aleiner96}.

Apart from the setting of weak localization, effects of a finite
Ehrenfest time have received much attention recently: The
excitation gap in an Andreev billiard~\cite{REF:ABgap} as well as
the shot noise~\cite{REF:shotnoise} of a ballistic cavity are
predicted to be suppressed when $\tau_{\rm E}>\tau_{\rm D}$. The
latter effect have received experimental
support~\cite{REF:ShNoiseexp}. For these problems there now exist
semiclassical theories, which do not invoke impurity scattering.
However, all these theories deal only with leading order effects.
Quantum corrections such as weak localization are beyond their
reach. That is why in this work we follow an altogether different
approach.

%------------------------------------
\begin{figure}
\begin{center}
\mbox{ \includegraphics[width=8cm]{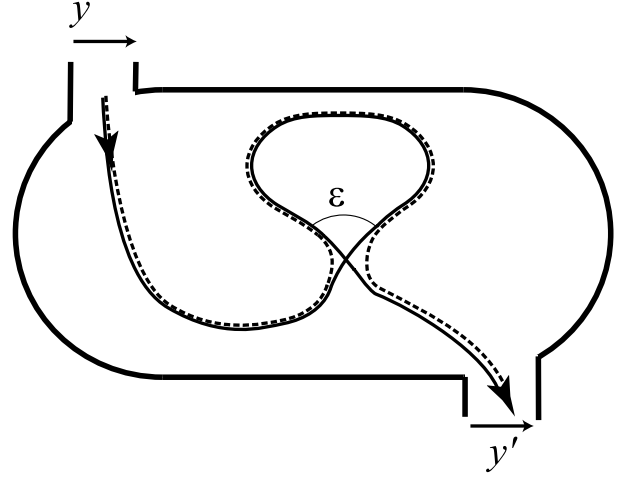}}
\caption{The Richter-Sieber pair. The weak localization correction to the transmission amplitudes comes from selfcrossing
angles $\epsilon\lesssim\sqrt{\lambda\hbar/E_F}$. The characteristic time of such orbits is the Ehrenfest time
$\tau_E=(1/\lambda)\ln(E_F/\lambda\hbar)$}.
\label{FIG:SRpair}
\end{center}
\end{figure}
%------------------------------------

Following  Richter and Sieber, we consider a two-dimensional ballistic quantum dot to which two leads
of width $w$ and $w'$ are attached. We assume that the classical
dynamics of this dot is chaotic, with Lyapunov exponent $\lambda$.
The Landauer formula for the conductance is given by
\begin{equation}
G
%(E,B)=2\frac{e^2}{h} {\mathcal T}
=2\frac{e^2}{h}\sum_{n=1}^{N}\sum_{m=1}^{N'}|t_{nm}|^2,
\end{equation}
where $t_{nm}$ is the transmission amplitude between incoming and
outgoing channels $m$ and $n$ at the Fermi energy $E_{F}$ and $N$($N'$) is the number of channels
of width $w$($w'$).
The
semiclassical expression for $t_{nm}$ is given as a sum over classical
trajectories $\gamma$ joining two leads~\cite{REF:Stone,REF:Sieber}:
\begin{equation}
t_{nm}=-\sqrt{\frac{\pi\hbar}{2 w w'}}
\sum_{\gamma(\bar{n},\bar{m})}\frac{\exp\big((i/\hbar)S_\gamma\big)\,\Phi_\gamma}
{|\cos\theta_{\bar{n}} \cos\theta_{\bar{m}} M^\gamma_{21}|^{1/2}}.
\end{equation}
Here $\sin\theta_{\bar{n}}=\bar{n}\pi/k_F w$ and $\sin\theta_{\bar{m}}=\bar{m}\pi/k_F w'$,
$\bar{n}=\pm n$ and $\bar{m}=\pm m$, and
$\Phi_\gamma={\rm sgn}(\bar{m}){\rm sgn}(\bar{n})\exp\big( i\pi(\bar{m}y/w-\bar{n}y'/w'-\mu_\gamma/2+1/4)\big)$ .
The term $S_\gamma$ is the classical action,
$M^\gamma_{21}$ is an element of the monodromy matrix, and $\mu_\gamma$ is the Maslov index.
The trajectory $\gamma$ starts at transverse coordinate $y$ in lead $w$ with an angle $\theta_{\bar{n}}$
and  ends at the transverse coordinate $y'$ in lead $w'$ with angle $\theta_{\bar{m}}$.

When calculating
$|t_{nm}|^2$ the double sum over trajectories $\gamma$ and $\gamma'$ is approximated to leading order by
the diagonal approximation $\gamma=\gamma'$~\cite{REF:Stone}.
The first order quantum correction to the transmission amplitudes (responsible for the
weak localization effect~\cite{REF:unitarity}) is due to Richter-Sieber pairs~\cite{REF:Sieber}:
$\gamma$ is exponentially close to $\gamma'$ everywhere
except in the vicinity of a crossing point of $\gamma$ where $\gamma'$ avoids that crossing. This is illustrated
in Fig.~1. The action
difference between $\gamma$ and $\gamma'$ is: $\Delta S=E_F \epsilon^2/\lambda$, where $\epsilon$
is the angle at the crossing. In the diagonal approximation, the sum over trajectories
can be evaluated via the sum rule~\cite{REF:Sieber}
\begin{equation}
\sum_{\gamma(y',\theta_n;y,\theta_m)}
\frac{\delta(T-T_\gamma)}{|M^{\gamma}_{21}|}
=\frac{ \cos\theta_n \cos\theta_m}{2\pi mA}
dy\,dy'\,\rho(T),
\label{EQ:sumrule}
\end{equation}
where the sum is over all trajectories that begin in interval $dy'$ around $y$ and
end in interval $dy$ around $y$,
$\rho(T)\propto\exp(-T/\tau_{\rm D})$ is the dwell time distribution
and
\begin{equation}
\tau_{\rm D}=mA/\hbar(N+N')
\label{EQ:esctime}
\end{equation}
is the mean dwell time, we denote by $m$ the effective electron mass, by $A$ the area of the cavity, and by
$N=k_{F}w/\pi$, $N'=k_{F}w'/\pi$ the number of channels in the two leads.
The weak localization correction from Richter-Sieber pairs
is given by
\begin{eqnarray}
\delta|t_{nm}|^2=\frac{2E_F \hbar}{\pi m^2 A^2}
\int_0^\pi d\epsilon \int_{T_\epsilon}^\infty dT
{\rm e}^{-T/\tau_{\rm D}}(T-T_\epsilon)^2 \nonumber \\
\times
\cos( E_F\epsilon^2/\lambda\hbar)\sin\epsilon,
\label{EQ:Tamp}
\end{eqnarray}
where $T_\epsilon=-(2/\lambda)\ln\epsilon$. The lower bound
in the integral over $T$
signifies that there are no orbits shorter than
$T_\epsilon$ with a selfcrossing angle $\epsilon$.

So far we have followed the calculation of
Richter and Sieber~\cite{REF:Sieber}. Now we depart from it. We first evaluate the $T$ integral,
\begin{equation}
\delta|t_{nm}|^2=\frac{4E_F\hbar\tau_{\rm D}^3}{\pi m^2 A^2}
\int_0^\pi d\epsilon \,{\rm e}^{-T_\epsilon/\tau_{\rm D}}
\cos (E_F\epsilon^2/\lambda\hbar)\,\sin\epsilon.
\end{equation}
In the semiclassical limit, the main contribution to this integral comes from
$\epsilon\lesssim\sqrt{\lambda\hbar/E_F}\ll 1$. Thus we may approximate
$\sin\epsilon\approx\epsilon$ and extend the upper limit of the integral to infinity.
The result is
\begin{eqnarray}
\delta|t_{nm}|^2&=&\frac{4E_F\hbar}{\pi m^2 A^2} \tau_{\rm D}^3
\int_0^\infty d\epsilon \, \epsilon^{1+2/\lambda\tau_{\rm D}} \cos({E_F\epsilon^2/\lambda\hbar})
\nonumber \\
&=&-\left(\frac{\hbar\tau_{\rm D}}{m A}\right)^2\frac{2\lambda\tau_{\rm D}}{\pi}
\sin\left(\frac{\pi}{2\lambda\tau_{\rm D}}\right)
\Gamma\left(1+\frac{1}{\lambda\tau_{\rm D}}\right)
\nonumber \\
&&\qquad\times \exp({-\tau_{\rm E}/\tau_{\rm D}}),
\end{eqnarray}
where $\tau_{\rm E}=(1/\lambda)\ln(E_F/\lambda\hbar)$ is the Ehrenfest time of this problem.
In the relevant regime $\lambda\tau_{\rm D}\gg 1$
we have
\begin{equation}
\delta|t_{nm}|^2\simeq \left(\frac{\hbar\tau_{\rm D}}{m A}\right)^2 {\rm e}^{-\tau_{\rm E}/\tau_{\rm D}}.
\end{equation}
Finally, using Eq.~(\ref{EQ:esctime}) and the sum rule~(\ref{EQ:sumrule}), we find the weak localization
correction to the conductance
\begin{equation}
\delta G_= -\frac{2e^2}{h}\frac{N N'}{(N+N')^2}\exp({-\tau_{\rm E}/\tau_{\rm D}}),
\end{equation}
in agreement with Ref.~\cite{REF:Aleiner96}.

Up to this point we have rederived a known result. Now we shall apply this technology to the magnetic field
dependence of the weak localization correction in the Ehrenfest regime. This is done via the calculation
of the magnetic field dependence of the density of self crossings~\cite{REF:Sieber}. Accordingly,
Eq.(\ref{EQ:Tamp}) is modified as follows:
\begin{eqnarray}
\delta|t_{nm}|^2&=&\frac{4E_F \hbar\tau_{\rm B}^2}{\pi m^2 A^2}
\int_0^\pi d\epsilon \int_{T_\epsilon}^\infty dT
\cos( E_F\epsilon^2/\lambda\hbar)\sin\epsilon
\nonumber \\
&&\times{\rm e}^{-T/\tau_{\rm D}} \left( {\rm e}^{(T_\epsilon-T)/\tau_{\rm B}}-1+\frac{T-T_\epsilon}{\tau_{\rm B}}\right),
\end{eqnarray}
where $\tau_{\rm B}=\phi_0^2/(8\pi^2\beta B^2)$ is the magnetic time, $\phi_0$ is the flux quantum, $B$ is the
magnetic field, and $\beta$ is a
system dependent parameter~\cite{REF:Stone,REF:Sieber}. As before, we first evaluate the $T$ integral exactly and then
evaluate the $\epsilon$ integral in stationary phase approximation. This produces the $B$ dependent transmission
matrix elements $\delta|t_{nm}(B)|^2=\delta|t_{nm}(0)|^2({1+\tau_{\rm D}/\tau_{\rm B}})^{-1}$. Finally, summing
over all channels we obtain the magnetic field dependence of the weak localization correction to the conductance,
\begin{equation}
\delta{G}(B)=-\frac{2e^2}{h}
\frac{N N'}{(N+N')^2}\frac{{\rm e}^{-\tau_{\rm E}/\tau_{\rm D}}}{1+\tau_{\rm D}/\tau_{\rm B}}
\end{equation}
We see that the Lorentzian lineshape of the weak localization
peak is preserved in the Ehrenfest regime, while its size is exponentially suppressed.

In conclusion, we have presented a derivation of the Ehrenfest time dependence of the weak localization correction
in a two dimensional chaotic billiard. All interference effects are fully accounted for within the framework of a
controlled semiclassical approximation~\cite{REF:Sieber}, without requiring the artificial inclusion of impurity
scattering~\cite{REF:Aleiner96}. Interesting extensions include the appearance of a second Lyapunov exponent in
three dimensions, and the coexistence of chaotic and mixed regions of phase space. It would also be
of interest to extend the method to describe universal conductance fluctuations in the Ehrenfest regime.

This work was supported by the Dutch Science Foundation NWO/FOM. We thank C.W.J. Beenakker and J. Tworzydlo for
helpful discussions.


\begin{thebibliography}{99}
\bibitem{REF:unitarity}
Equivalently, one can calculate the quantum correction to the reflection amplitudes, but this is
more complicated because then both diagonal terms and Richter-Sieber pairs contribute~\cite{REF:Sieber}.
\end{thebibliography}
\end{document}